\documentclass[aps,prd,superscriptaddress,showpacs,nofootinbib,notitlepage,onecolumn]{revtex4-1}
\usepackage{graphicx,subfigure,amsmath,amsfonts,amssymb,multirow,mathrsfs}
\usepackage[colorlinks,linkcolor=blue,citecolor=blue,urlcolor=blue]{hyperref}

\def\apj{ApJ~}                 
\def\apjl{ApJL~}                
\def\apjs{ApJS~}               
\def\aap{A\&A~}                

\def\cjaa{Chinese J Astron Astrophys~}

\def\mnras{MNRAS~}             
\def\prc{Phys~Rev~C~}        
\def\prd{Phys~Rev~D~}        
\def\prl{Phys~Rev~Lett~}    
\def\nat{Nature~}              

\begin{document}

\title{Mass and radius of the most massive neutron star: The probe of the equation of state and perturbative QCD}

\author{Shao-Peng Tang}
\affiliation{Key Laboratory of Dark Matter and Space Astronomy, Purple Mountain Observatory, Chinese Academy of Sciences, Nanjing 210033, China}
\author{Ming-Zhe Han}
\affiliation{Key Laboratory of Dark Matter and Space Astronomy, Purple Mountain Observatory, Chinese Academy of Sciences, Nanjing 210033, China}
\author{Yong-Jia Huang}
\affiliation{Key Laboratory of Dark Matter and Space Astronomy, Purple Mountain Observatory, Chinese Academy of Sciences, Nanjing 210033, China}
\affiliation{RIKEN Interdisciplinary Theoretical and Mathematical Sciences Program (iTHEMS), RIKEN, Wako 351-0198, Japan}
\author{Yi-Zhong Fan}
\email[Corresponding author.~]{yzfan@pmo.ac.cn}
\affiliation{Key Laboratory of Dark Matter and Space Astronomy, Purple Mountain Observatory, Chinese Academy of Sciences, Nanjing 210033, China}
\affiliation{School of Astronomy and Space Science, University of Science and Technology of China, Hefei, Anhui 230026, China}
\author{Da-Ming Wei}
\affiliation{Key Laboratory of Dark Matter and Space Astronomy, Purple Mountain Observatory, Chinese Academy of Sciences, Nanjing 210033, China}
\affiliation{School of Astronomy and Space Science, University of Science and Technology of China, Hefei, Anhui 230026, China}
\date{\today}

\begin{abstract}
Recently, an association of GW190425 and FRB 20190425A had been claimed and a highly magnetized neutron star (NS) remnant was speculated.
Given the $\sim 2.5$-h delay of the occurrence of FRB 20190425A, a uniformly rotating supramassive magnetar is favored since the differential rotation would have been promptly terminated by the magnetic braking.
The required maximum gravitational mass ($M_{\rm TOV}$) of the nonrotating NS is $\approx 2.77M_\odot$, which is strongly in tension with the relatively low $M_{\rm TOV}\approx 2.25M_\odot$ obtained in current equation of state (EOS) constraints incorporating perturbative quantum chromodynamics (pQCD) information.
However, the current mass-radius and mass-tidal deformability measurements of NSs alone do not convincingly exclude the high $M_{\rm TOV}$ possibility.
By performing EOS constraints with mock measurements, we find that with a $2\%$ determination for the radius of PSR J0740+6620-like NS it is possible to distinguish between the low and high $M_{\rm TOV}$ scenarios.
We further explore the prospect to resolve the issue of the appropriate density to impose the pQCD constraints with future massive NS observations or determinations of $M_{\rm TOV}$ and/or $R_{\rm TOV}$.
It turns out that measuring the radius of a PSR J0740+6620-like NS is insufficient to probe the EOSs around 5 nuclear saturation density, where the information from pQCD becomes relevant.
The additional precise $M_{\rm TOV}$ measurements anyhow could provide insights into the EOS at such a density.
Indeed, supposing the central engine of GRB 170817A is a black hole formed via the collapse of a supramassive NS, the resulting $M_{\rm TOV}\approx 2.2M_\odot$ considerably softens the EOS at the center of the most massive NS, which is in favor of imposing the pQCD constraint at density beyond the one achievable in the NSs.
\end{abstract}
\maketitle

\section{Introduction} \label{sec:intro}
The maximum mass of a nonrotating neutron star (NS), denoted as $M_{\rm TOV}$, is a crucial parameter in studying the equation of state (EOS) of cold dense matter.
PSR J0740+6620 ($2.08\pm0.07M_\odot$ \citep{2021ApJ...915L..12F}) is recognized as the most massive NS observed to date, whose mass can serve as the lower limit of $M_{\rm TOV}$.
Recently, a mass estimate for the black-widow binary pulsar PSR J0952-0607 ($M=2.35\pm0.17M_\odot$ \citep{2022ApJ...934L..17R}; note that the rapid rotation have enhanced the gravitational mass by $\sim 0.07M_\odot$), albeit with significant uncertainty, appears to update this record.
Other investigations have also calculated the $M_{\rm TOV}$ from the mass distribution of NSs \citep{2018MNRAS.478.1377A, 2020PhRvD.102f3006S, Fan:2023spm} and the observations of short gamma-ray bursts \citep{2013PhRvD..88f7304F} in particular the multimessenger observations of GW170817/GRB 170817A \citep{2017ApJ...850L..19M, 2017PhRvD..96l3012S, 2018ApJ...852L..25R, 2018PhRvD..97b1501R, 2020PhRvD.101f3029S, 2020ApJ...904..119F}.
The detection of GW190814 has also prompted some studies to constrain the lower limit of $M_{\rm TOV}$ \citep{2020PhRvC.102f5805F, 2020MNRAS.499L..82M, 2020ApJ...902...38Z, 2020ApJ...904...39H, 2021ApJ...908L...1T, 2021ApJ...908..122G, 2021ApJ...908L..28N, 2021ApJ...910...62Z}, but this heavily depends on the assumptions regarding the nature and the spin of the secondary component \citep{2020MNRAS.499L..82M, 2021ApJ...908L..28N}.

GW190425 \citep{2020ApJ...892L...3A} is widely believed to be the second identified binary neutron star (BNS) merger gravitational wave (GW) event (though the possibility of a neutron star$\textendash$very light black hole merger cannot be ruled out, as shown in Ref.~\citep{2020ApJ...891L...5H}).
This binary system has a total gravitational mass substantially larger than those observed in Galactic double neutron star systems \citep{2019ApJ...876...18F}.
Recently, \citet{2023NatAs...7..579M} claimed that GW190425 coincided with a bright, nonrepeating fast radio burst (FRB) event, FRB 20190425A \citep{2021ApJS..257...59C}, which took place 2.5 hours after the GW event.
The magnetar origin of FRBs is supported by current observations (e.g.,  Refs.~\citep{2020Natur.587...59B, 2018Natur.553..182M, 2020Natur.587...54C, 2020ApJ...898L..29M, 2021NatAs...5..378L}; see also \citet{2002ApJ...580L..65L} for the pioneering prediction of FRBs from magnetars).
If the coincidence between these two transient events is held (see also e.g., Refs.~\citep{2023arXiv230600948B, 2024MNRAS.528.5836R, 2024MNRAS.528.2299S} for doubt on this accuracy) and given the 2.5-h delay, FRB 20190425A should be from a uniformly rotating supermassive NS (SMNS) rather than a hypermassive NS, since the magnetic field could brake the differential rotation in $\sim 0.1~{\rm s}~(B_{\rm s}/10^{15}~{\rm Gauss})^{-1}$ timescale \citep{2000ApJ...544..397S,2006ChJAA...6..513G}, where $B_{\rm s}$ is the dipole magnetic field strength of the magnetar.
Consequently, the formation of an SMNS would suggest a very large maximum mass for a nonrotating NS.

On the other hand, the radii of massive NSs (including the radius of the most massive one, i.e., $R_{\rm TOV}$) are key to probing high-density EOSs \citep{2020PhRvD.101l3007L, 2020ApJ...899..164H}.
Direct radius measurements of massive NSs can be conducted by x-ray timing observations if we can find some suitable massive candidates \citep{2019ApJ...887L..27G, 2021ApJ...918L..27R, 2021ApJ...918L..28M}.
Another indirect method involves estimating the tidal deformability or radius by using the quasi-universal relation (derived from numerical simulations) with post-merger GW frequency \citep{2019PhRvD.100j4029B, 2022PhRvL.128p1102B}.
Recently, \citet{2024ApJ...960...67T} suggested that by measuring the masses and spins of a series of SMNS-originated black holes (BHs) through the mergers of second-generation (2G) BHs with NSs, the $M_{\rm TOV}$ and $R_{\rm TOV}$ could be determined simultaneously.
\citet{Barr:2024wwl} reported that PSR J0514-4002E, a millisecond pulsar in an eccentric binary system with a total mass of $3.887 \pm 0.004M_\odot$ and situated in the globular cluster NGC 1851, is likely to have a 2G BH as its companion.
This finding lends further support to the scenario proposed by \citet{2024ApJ...960...67T}.
Some studies on universal relations have found potential correlations between the macroscopic (e.g., $M_{\rm TOV}$ and $R_{\rm TOV}$) and microscopic (e.g., the central pressure $p_c$ and density $n_c$) properties of the maximum-mass nonrotating NS \citep{2020PhRvD.101j3029O, 2023ApJ...949...11J, 2023ApJ...952..147C}.
Measurements of the mass and radius of the most massive nonrotating NS enable us to explore regions of higher density that remain inaccessible for a $1.4M_\odot$ NS.
Recent progress in {\it ab initio} quantum chromodynamics (QCD) computations has attracted considerable interest for their contribution to establishing the EOS for NS \citep{2022PhRvL.128t2701K, 2022arXiv221111414K,  2023ApJ...950..107G, Fan:2023spm, 2023SciBu..68..913H, 2023PhRvD.108i4014B, 2023PhRvD.107a4011B, 2023arXiv230902345M, 2023PhRvC.107e2801S, 2023arXiv230711125Z}.
However, there remains ongoing debate about the density at which perturbative QCD (pQCD) constraints should be implemented (see e.g., Refs.~\citep{2023PhRvC.107e2801S, 2023ApJ...950..107G, 2023PhRvD.108d3013E, 2023arXiv230711125Z}).

In this work, we first evaluate whether the coincidence between GW190425 and FRB 20190425A aligns with current EOS constraints, specifically those that include pQCD information.
We find that there is a strong tension between such scenario and the pQCD constraints.
Though the Bayesian evidence for the result obtained with pQCD constraints is marginally higher, we are currently unable to definitively dismiss the GW190425/FRB 20190425A association.
We then aim to determine whether future radius measurements of massive NSs can distinguish between the low- and high-$M_{\rm TOV}$ scenarios.
By performing EOS constraints with mock measurements, we find that with a $2\%$ uncertainty for the radius of a PSR J0740+6620-like NS, it is possible to definitively determine whether the GW190425/FRB 20190425A association is real or not.
We further investigate whether we can resolve the issue of the appropriate density at which to impose the pQCD constraints with future massive NS observations or determinations of $M_{\rm TOV}$ and $R_{\rm TOV}$.
We find that measuring the radius of a PSR J0740+6620-like NS is insufficient to probe the EOSs at the density where pQCD takes effect.
However, $M_{\rm TOV}$ and $R_{\rm TOV}$ measurements could determine whether the pQCD constraints imposed at $10n_s$ are reliable or not.

\section{Methods}\label{sec:methods}
There are two types of models for phenomenologically constructing the EOS.
The first type includes parametrization methods such as the piecewise polytropes \citep{2009PhRvD..79l4032R} and spectral decomposition \citep{2014PhRvD..89f4003L}.
The second type encompasses nonparametric methods like the Feed-Forward-Neural-Network (FFNN) method \citep{2021ApJ...919...11H, 2023ApJ...950...77H} and the Gaussian Process (GP) method \citep{2019PhRvD..99h4049L, 2020PhRvD.101f3007E, 2020PhRvD.101l3007L}.
Due to its flexibility, the GP regression has been extensively utilized. 
In this study, we employ the methodology of \citet{2023ApJ...950..107G} and \citet{Fan:2023spm} to implement the GP in generating an EOS ensemble.
We connect the low-density (below approximately $0.3n_s$, where $n_s$ is the nuclear saturation density) part of SLy \citep{2001A&A...380..151D} to the randomly selected EOS from chiral effective field theory ($\chi$EFT) results (i.e., 1000 EOSs from the publicly available dataset provided by \citet{2019PhRvL.122d2501D}) between the densities of $0.3n_s$-$1n_s$.
Above $1n_s$, we apply the GP and condition it between $1n_s$ and $1.1n_s$.
Before conditioning, $\phi(n)=-\ln\left(1/c_s^2(n)-1\right)$, an auxiliary variable related to the speed of sound, is assumed to follow a normal distribution, $\phi(n) \sim \mathcal{N}\left( -\ln(1/\bar{c}_s^2-1), K(n,n^{\prime}) \right)$, where $K(n,n^{\prime})=\eta e^{-(n-n^{\prime})^2/2l^2}$ is the Gaussian kernel.
The distributions of the three hyperparameters, namely the variance $\eta$, the correlation length $l$, and the mean speed of sound squared $\bar{c}_s^2$ are $\eta \sim \mathcal{N}(1.25,0.2^2)$, $l \sim \mathcal{N}\left(0.5n_s, (0.25n_s)^2\right)$, and $\bar{c}_s^2 \sim \mathcal{N}(0.5,0.25^2)$, respectively.
The conditioning is performed according to the following formulas,
\begin{equation}
\begin{aligned}
&\phi_{\rm GP}^{*} \mid n_{\rm CET}, \bar{\phi}_{\rm CET}, \sigma^2_{\phi_{\rm CET}}, n_{\rm GP} ~\sim~ \mathcal{N}\left(\bar{\phi}_{\rm GP}^{*}, {\rm cov}(\phi_{\rm GP}^{*})\right), \\
&\bar{\phi}_{\rm GP}^{*} = \bar{\phi}_{\rm GP} + K(n_{\rm GP}, n_{\rm CET})[K(n_{\rm CET}, n_{\rm CET})+4\sigma^2_{\phi_{\rm CET}}I]^{-1}(\bar{\phi}_{\rm CET}-\bar{\phi}_{\rm GP}), \\
&{\rm cov}(\phi_{\rm GP}^{*}) = K(n_{\rm GP}, n_{\rm GP})-K(n_{\rm GP}, n_{\rm CET})[K(n_{\rm CET}, n_{\rm CET})+4\sigma^2_{\phi_{\rm CET}}I]^{-1}K(n_{\rm CET}, n_{\rm GP}).
\end{aligned}
\end{equation}
where $\bar{\phi}_{\rm CET}$, $\sigma^2_{\phi_{\rm CET}}$ are the mean and variance of $\chi$EFT results at densities of $n_{\rm CET}$, $\bar{\phi}_{\rm GP}=-\ln(1/\bar{c}_s^2-1)$ , and $I$ is the identity matrix.
Please note that both $K(n_{\rm GP}, n_{\rm CET})$ and $K(n_{\rm CET}, n_{\rm CET})$ are Gaussian kernels. Their coefficients are consistent with the hyperparameters sampled for the GP and keep the same before/after conditioning.
We employ the Cholesky decomposition method when computing the inverse of $[K(n_{\rm CET}, n_{\rm CET})+4\sigma^2_{\phi_{\rm CET}}I]$ to guarantee accuracy.
In our work, the purpose of conditioning the GP is to ensure a smooth transition from the GP to the $\chi$EFT EOSs ($0.3n_s$-$1n_s$ part). For the hyperparameters of the covariance function $K(n,n^{\prime})$, we sample them from their respective hyperprior distributions once for each EOS sample. This means that each time we generate a new EOS sample using the GP, we employ a fresh set of hyperparameters sampled from the hyperpriors.
In practice, we draw a sample of $\phi_{\rm GP}^{*}(n)$ from the conditioned GP and solve for the speed of sound $c_s$, the baryon chemical potential $\mu$, the energy density $\varepsilon$, and the pressure $p$ by using 
\begin{equation}
\begin{aligned}
&c_s^2(n) = 1/(e^{-\phi_{\rm GP}^{*}(n)}+1), \\
&\mu(n) = \mu_0(n) \exp\!\!\left( \int_{n_0}^n dn^{\prime} c_s^2(n^{\prime})/n^{\prime} \right), \\
&\varepsilon(n) = \varepsilon_0 + \int_{n_0}^n dn^{\prime} \mu(n^{\prime}),~p(n) = -\varepsilon(n)+\mu(n)n.
\end{aligned}
\end{equation}
With the nonparametric EOSs in hand, we then construct their corresponding mass-radius ($M-R$) and mass-tidal deformability ($M-\Lambda$) curves by solving the Tolman-Oppenhimer-Volkoff and Regge-Wheeler equations \citep{2014PhRvD..89f4003L}.

The likelihoods of the NICER observations (i.e., the mass and radius measurements for PSR J0030+0451 and PSR J0740+6620), the tidal deformability measurements of GW170817, and the pQCD constraints are identical to those used in \citet{2023ApJ...950..107G} and \citet{Fan:2023spm}.
The likelihood of forming an SMNS remnant from GW190425 under a specified EOS is evaluated as
\begin{equation}
p({\rm SMNS}\mid \boldsymbol{\theta}^i,{\rm EOS}) = \frac{1}{N}\sum_i^{N} \mathcal{H}(M_{\rm b,1}^i+M_{\rm b,2}^i<M_{\rm b,crit}),
\end{equation}
where $M_{\rm b,1}^i$, $M_{\rm b,2}^i$ are the baryonic component masses of GW190425 derived from the low-spin posteriors $\boldsymbol{\theta}^i$ \citep{2020ApJ...892L...3A}, $\mathcal{H}$ is the Heaviside step function, and $M_{\rm b,crit} = M_{\rm crit}(1 + {\rm BE}_{\rm crit}/M_{\rm crit})$ is the critical baryonic mass that a uniformly rotating SMNS can support.
The critical gravitational mass $M_{\rm crit}$ and the corresponding reduced binding energy ${\rm BE}_{\rm crit}/M_{\rm crit}$ fulfill the following universal relations:
\begin{equation}
\begin{aligned}
&M_{\rm crit} = (1+0.0902\mathscr{C}_{\rm TOV}^{-1}\chi_{\rm coll}^2+0.0193\mathscr{C}_{\rm TOV}^{-2}\chi_{\rm coll}^4)M_{\rm TOV},\\
&\frac{{\rm BE}_{\rm crit}}{M_{\rm crit}} = -0.10+0.78(1-0.050\chi_{\rm coll}-0.034\chi_{\rm coll}^2)\mathscr{C}_{\rm TOV}\\
&\qquad\qquad+0.61(1+0.23\chi_{\rm coll}-0.58\chi_{\rm coll}^2)\mathscr{C}_{\rm TOV}^2,
\end{aligned}
\end{equation}
where $\mathscr{C}_{\rm TOV}$ is the compactness of the maximum-mass nonrotating NS and $\chi_{\rm coll}$ is the dimensionless angular momentum at the time of collapse \citep{2020PhRvD.101f3029S}.
These universal relations were first introduced by \citet{2016MNRAS.459..646B} and later refined by \citet{2020PhRvD.101f3029S} (see also Ref.~\citep{2020MNRAS.496L..16M}). They were obtained using the rotating relativistic NS model, implemented in the rotating neutron star (RNS) code \citep{1994ApJ...422..227C, 1995ApJ...444..306S, 1996PhDT.........4S}. This model is appropriate for the scenario involving FRB 20190425A, which exhibited a 2.5-h delay relative to the occurrence of GW190425. Given the significant cooling of the remnant NS over this duration, and cessation of differential rotation, employing a uniformly rotating model without thermal effects is a reasonable approximation. Furthermore, the robustness of these relations have been tested against a wide range of EOSs with high accuracy.
We set $\chi_{\rm coll}=0.65$, which is close to the maximum dimensionless spin of uniformly rotating NSs \citep{2020MNRAS.499L..82M}.
Please note that a smaller $\chi_{\rm coll}$ will result in a larger $M_{\rm TOV}$, leading to even more significant tension (refer to Sec. III, Results). 
We also examine the EOS constraints that future precise mass-radius measurements of the massive NS PSR J0740+6620, or the maximum-mass nonrotating NS, could impose.
We presume a hypothetical mass-radius measurement of PSR J0740+6620, with $R_{\rm J0740}=12.44\pm0.25{\rm km}$ (corresponding to a $2\%$ relative uncertainty) and a fixed mass $M_{\rm J0740}=2.08M_{\odot}$ (referred to as `Mock ($M_{\rm J0740}$, $R_{\rm J0740}$)').
The hypothetical mass-radius measurement of the maximum-mass nonrotating NS is assumed to be $M_{\rm TOV}=2.25\pm0.01M_{\odot}$ and $R_{\rm TOV}=11.9\pm0.6{\rm km}$ (referred to as `Mock ($M_{\rm TOV}$, $R_{\rm TOV}$)'), with uncertainties based on prospects presented in \citet{2024ApJ...960...67T}.
Although these prospects are promising, we must wait for the construction of the third-generation ground-based GW detectors.
We then explore how a precision of $\pm 0.04M_\odot$ for $M_{\rm TOV}$ (which may be achievable from the multimessenger observations of future detectable BNS mergers) in conjunction with the mass-radius measurement of PSR J0740+6620 (referred to as `Mock ($M_{\rm TOV}$, $R_{\rm J0740}$)') could constrain the EOS.
All of the hypothetical median values above are assigned based on the results obtained with the `Data+p($M_{\rm max}$)+pQCD($10n_s$)' dataset in \citet{Fan:2023spm}, and the uncertainties are given for a 1-$\sigma$ confidence level.

\section{Results} \label{sec:results}
\subsection{How does the GW190425/FRB 20190425A association impact the EOSs?}
\begin{figure}
    \centering
    \includegraphics[width=0.49\textwidth]{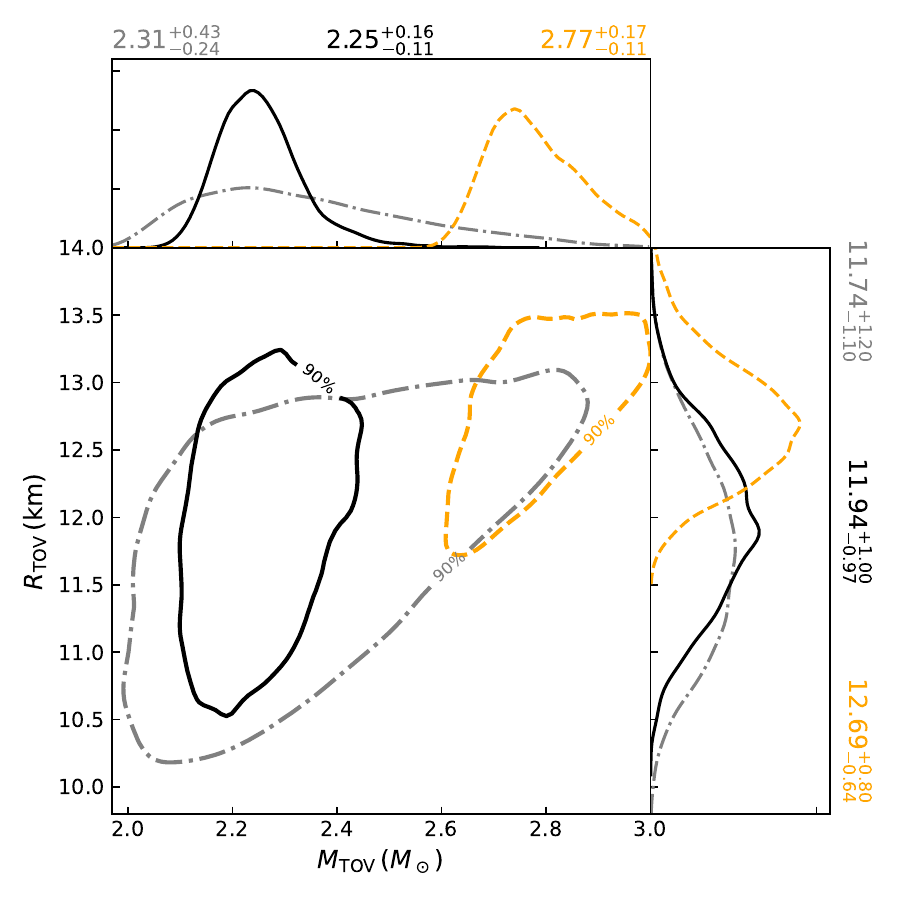}
    \includegraphics[width=0.49\textwidth]{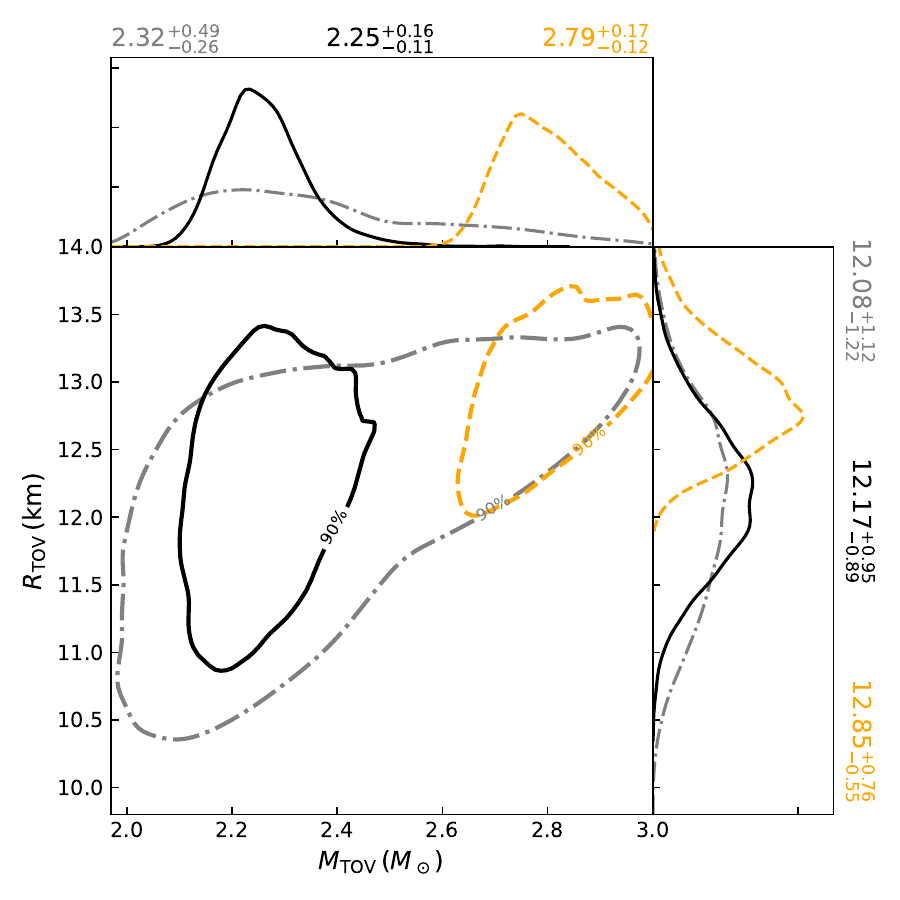}
    \caption{Posterior distributions of $M_{\rm TOV}$ and $R_{\rm TOV}$. The gray dash-dotted, black solid, and orange dashed lines represent the results derived with the fiducial `Data', `Data+p($M_{\rm max}$)+pQCD($10n_s$)' (adopted from Ref.~\citep{Fan:2023spm}), and `Data+FRB+SMNS' datasets, respectively. The primary distinction between the left and right panels lies in the source of the measurements: for both PSR J0030+0451 and PSR J0740+6620, the left panel utilizes data from Riley {\it et al.} \citep{2019ApJ...887L..21R, 2021ApJ...918L..27R}, whereas the right panel incorporates measurements from Miller {\it et al.} \citep{2019ApJ...887L..24M, 2021ApJ...918L..28M}.}
    \label{fig:mtov-rtov}
    \hfill
\end{figure}
We present the reconstructed mass ($M_{\rm TOV}$) and radius ($R_{\rm TOV}$) of the maximum-mass nonrotating NS in Fig.~\ref{fig:mtov-rtov}.
The gray dash-dotted lines represent the results based on the fiducial dataset (hereafter referred to as `Data'): the $\chi$EFT calculations \citep{2019PhRvL.122d2501D}, the NICER observations \citep{2019ApJ...887L..21R,2019ApJ...887L..24M,2021ApJ...918L..27R,2021ApJ...918L..28M} (here we employ the results of Riley {\it et al.} as the default for both PSR J0030+0451 and PSR J0740+6620 and employing the findings from Miller {\it et al.} does not significantly modify our main results), and the tidal deformability measurements from GW170817 \citep{2019PhRvX...9a1001A}.
The black solid lines, on the other hand, depict the results that incorporate additional pQCD constraints imposed at $10n_s$ (i.e., the results obtained with the `Data+p($M_{\rm max}$)+pQCD($10n_s$)' dataset in Ref.~\citep{Fan:2023spm}).
If the potential association between GW190425 and FRB 190425A is established, the BNS merger remnant would undergo an SMNS phase required to produce the FRB as suggested by \citet{2023NatAs...7..579M}.
The orange dashed lines, therefore, represent the results for incorporating the $p({\rm SMNS}\mid \boldsymbol{\theta}^i,{\rm EOS})$ likelihood in addition to `Data' (hereafter referred to as `Data+FRB+SMNS').
We find that the assumption of forming an SMNS for GW190425 leads to a much heavier $M_{\rm TOV}$, which is in tension with that obtained with pQCD constraints.
The inferred $M_{\rm TOV}$ ($R_{\rm TOV}$) for the `Data+p($M_{\rm max}$)+pQCD($10n_s$)' and `Data+FRB+SMNS' results are $2.25_{-0.11}^{+0.16}M_{\odot}$ ($11.94_{-0.97}^{+1.00}\,{\rm km}$) and $2.77_{-0.11}^{+0.17}M_{\odot}$ ($12.69_{-0.64}^{+0.80}\,{\rm km}$), respectively.
Unless otherwise stated, all uncertainties are given at $90\%$ credibility.

\begin{figure}
    \centering
    \includegraphics[width=0.49\textwidth]{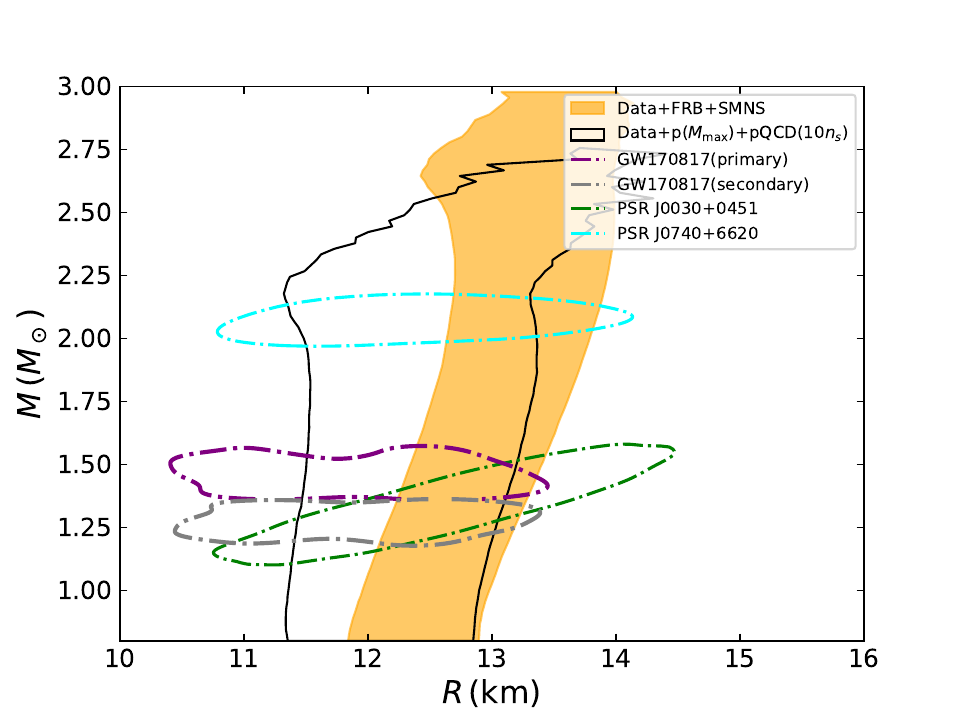}
    \includegraphics[width=0.49\textwidth]{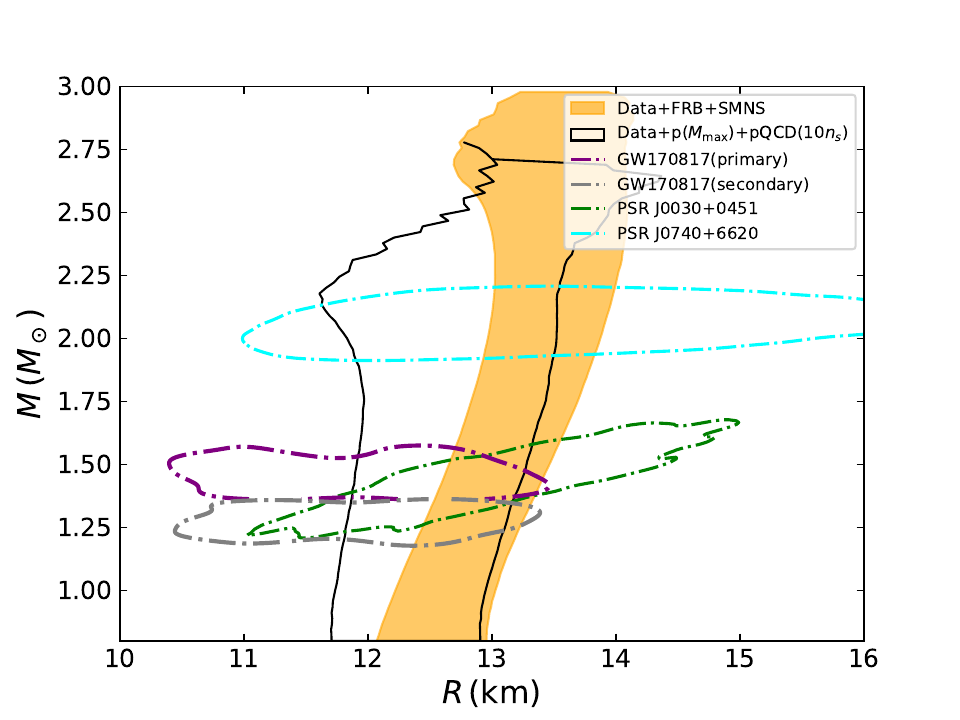}
    \caption{Reconstructed $90\%$ intervals of mass-radius curves for the `Data+p($M_{\rm max}$)+pQCD($10n_s$)' (black line; adopted from Ref.~\citep{Fan:2023spm}), and `Data+FRB+SMNS' (filled in orange) datasets.
The green, cyan, purple, and gray dash-dotted contours represent the $68.3\%$ mass-radius measurements for the isolated NS PSR J0030+0451, the massive NS PSR J0740+6620, the primary NS of GW170817, and the secondary component of GW170817 \citep{2018PhRvL.121p1101A}, respectively. The sole distinction between the left and right panels is the source of mass-radius measurements for PSR J0030+0451 and PSR J0740+6620. In the left panel, we incorporate data from Riley {\it et al.} \citep{2019ApJ...887L..21R, 2021ApJ...918L..27R}, whereas in the right panel, measurements are taken from Miller {\it et al.} \citep{2019ApJ...887L..24M, 2021ApJ...918L..28M}.}
    \label{fig:mr-range}
    \hfill
\end{figure}
As illustrated in Fig.~\ref{fig:mr-range}, the $90\%$ mass-radius intervals of the `Data+FRB+SMNS' are systematically shifted towards larger radii compared to those of the `Data+p($M_{\rm max}$)+pQCD($10n_s$)'.
\begin{figure}
    \centering
    \includegraphics[width=0.8\textwidth]{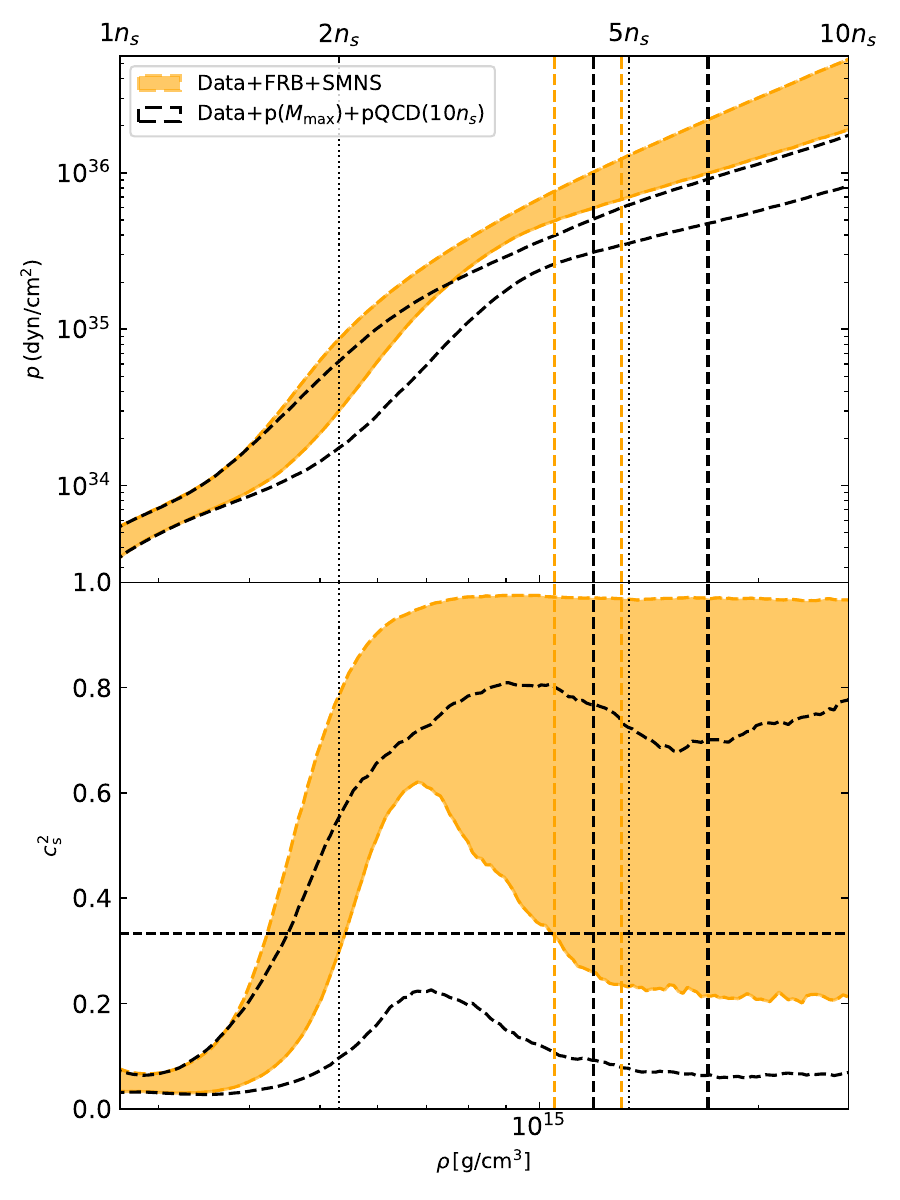}
    \caption{The $90\%$ pressure vs rest-mass density intervals (top panel) and the speed of sound squared vs rest-mass density intervals (bottom panel).
    The orange areas and black dashed lines represent the results of `Data+FRB+SMNS' and `Data+p($M_{\rm max}$)+pQCD($10n_s$)' (adopted from Ref.~\citep{Fan:2023spm}) datasets, respectively.
    The vertical dash-dotted lines mark the intervals of the central densities of the maximum-mass nonrotating NS.
    The horizontal black dashed line denotes the conformal limit.}
    \label{fig:eos-range}
    \hfill
\end{figure}
The EOSs, constrained by the `Data+FRB+SMNS' dataset, are significantly stiffer than those constrained by the `Data+p($M_{\rm max}$)+pQCD($10n_s$)' dataset, as clearly seen in Fig.~\ref{fig:eos-range}.
Furthermore, the $90\%$ $p-\rho$ ranges of the two results are in substantial conflict with each other.
The speed of sound squared shows a rapid increase for the `Data+FRB+SMNS' dataset, and the $90\%$ lower limit of $c_s^2$ peak is larger than approximately $0.6$, well above the so-called conformal limit.
The central density of the maximum-mass nonrotating NS of `Data+FRB+SMNS' ($n_{\rm c,TOV}=0.70_{-0.08}^{+0.07}\,{\rm fm}^{-3}$) is relatively smaller than that of `Data+p($M_{\rm max}$)+pQCD($10n_s$)' ($n_{\rm c,TOV}=0.87_{-0.15}^{+0.16}\,{\rm fm}^{-3}$), which is consistent with the correlation that a larger $R_{\rm TOV}$ predicts a smaller $n_{\rm c,TOV}$ \citep{2023ApJ...949...11J}.

\subsection{How do future $M-R$ measurements of massive NSs probe the high-density EOSs?}
\begin{figure}
    \centering
    \includegraphics[width=0.8\textwidth]{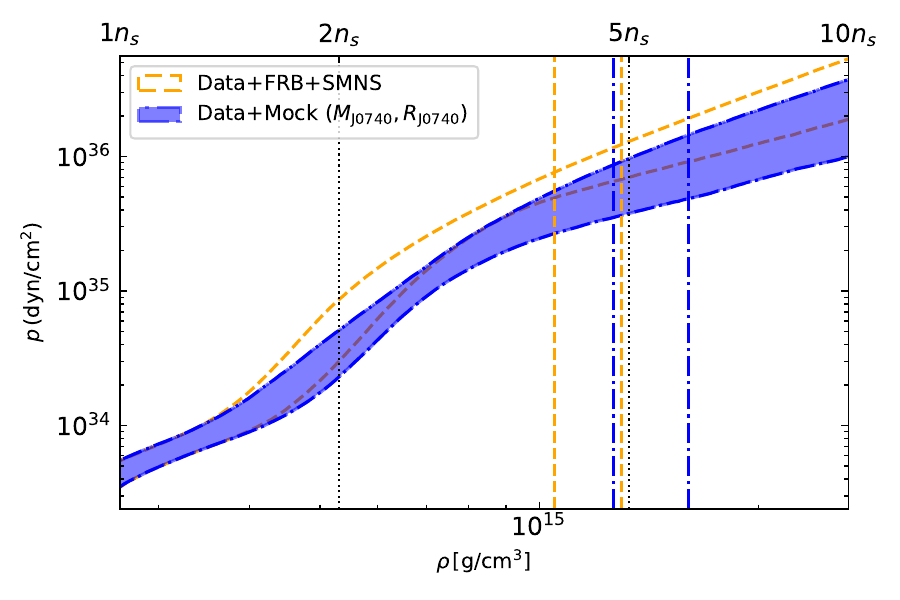}
    \includegraphics[width=0.8\textwidth]{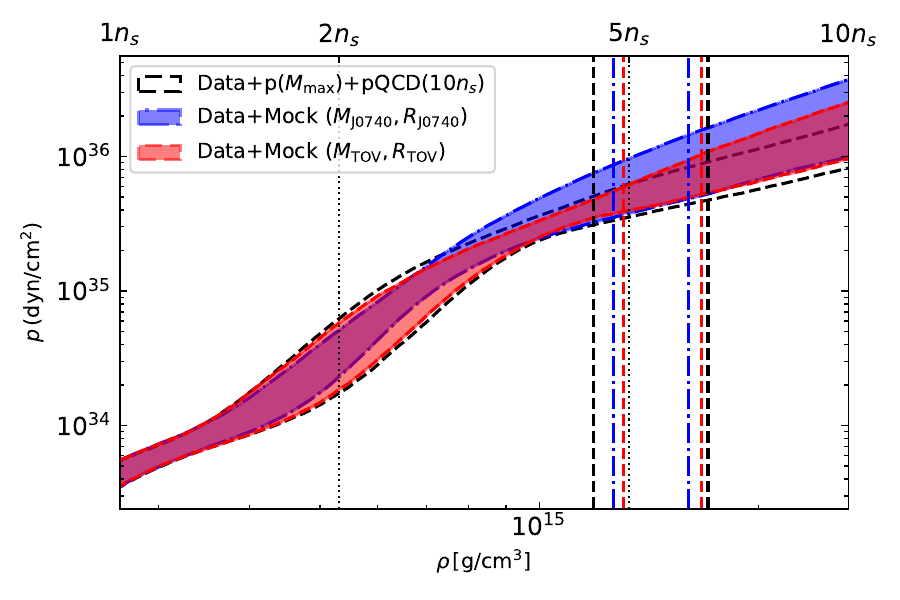}
    \caption{The $90\%$ pressure vs rest-mass density intervals.
    The areas filled in red and blue represent the results of `Data+Mock ($M_{\rm TOV}$, $R_{\rm TOV}$)'  and `Data+Mock ($M_{\rm J0740}$, $R_{\rm J0740}$)', respectively.
    The orange dashed lines in the top panel represent the results of the `Data+FRB+SMNS' dataset, which are identical to that presented in Fig.~\ref{fig:eos-range}.
    The black dashed lines in the bottom panel represent the results obtained with the `Data+p($M_{\rm max}$)+pQCD($10n_s$)' dataset as found in \citet{Fan:2023spm}.
    The vertical lines indicate the intervals of the central density of the maximum-mass nonrotating NS.}
    \label{fig:eos-range-mock1}
    \hfill
\end{figure}

The logarithmic Bayesian evidence of `Data+p($M_{\rm max}$)+pQCD($10n_s$)' is marginally greater than that of `Data+FRB +SMNS', with $\Delta \ln \mathcal{Z}\simeq1.6$ (if we adopt the mass-radius measurements from Miller {\it et al.} for both PSR J0030+0451 and PSR J0740+6620, $\Delta \ln \mathcal{Z}$ will reduce to $\simeq1.3$).
These Bayes factors do not definitively establish the correctness of either.
To determine whether future $M-R$ measurements of massive NSs can differentiate between the two, we perform EOS constraints with mock $M-R$ measurements of PSR J0740+6620 as outlined in Sec. II, Methods.
As illustrated in the top panel of Fig.~\ref{fig:eos-range-mock1}, we find that the EOSs within the density range, where the $p-\rho$ intervals of `Data+FRB+SMNS' and `Data+p($M_{\rm max}$)+pQCD($10n_s$)' diverge, can be constrained by the `Data+Mock ($M_{\rm J0740}$, $R_{\rm J0740}$)' dataset (please note that the `Data' here exclude the current measurements for PSR J0740+6620).
And the Bayes factor between `Data+Mock ($M_{\rm J0740}$, $R_{\rm J0740}$)' and `Data+FRB+SMNS' datasets is $\mathcal{B}\simeq38.0$ (i.e., $\log_{10} \mathcal{B} \simeq 1.6$).
As pointed out by \citet{jeffreys1961theory}, $\log_{10} \mathcal{B}> 1$ ($\log_{10} \mathcal{B} > 1.5$) can be interpreted as a strong (very strong) preference for one model over another, and $\log_{10} \mathcal{B} > 2$ as decisive evidence.
This implies that determining $R_{\rm J0740}$ with a relative uncertainty of $2\%$ makes it possible to distinguish between them.
Recently, \citet{2024ApJ...960...67T} proposed a novel method to precisely measure the mass and radius of the maximum-mass nonrotating NS.
We then proceeded to compare these measurements in terms of their ability to constrain the high-density EOSs with the constraining ability of $M-R$ measurements of massive NSs such as PSR J0740+6620.
As depicted in the bottom panel of Fig.~\ref{fig:eos-range-mock1}, we found that the `Data+Mock ($M_{\rm J0740}$, $R_{\rm J0740}$)' dataset only effectively constrains the EOSs below approximately $3n_s$.
In contrast, the `Data+Mock ($M_{\rm TOV}$, $R_{\rm TOV}$)' demonstrates a strong ability to constrain the EOSs at a density close to $n_{\rm c,TOV}$, compared to those obtained with mock $M-R$ measurements of PSR J0740+6620.
This suggests that even with very precise $M-R$ measurements for NSs, if the NSs were not heavy enough, it might not be possible to probe the EOSs at a density as high as the central density of a maximum-mass nonrotating NS.
It is reasonable to note that the two sets of EOSs, i.e., `Data+Mock ($M_{\rm TOV}$, $R_{\rm TOV}$)' and `Data+p($M_{\rm max}$)+pQCD($10n_s$)',  begin to show divergence beyond $n_{\rm c,TOV}$.
This divergence arises because the `Mock ($M_{\rm TOV}$, $R_{\rm TOV}$)' only contains the information for EOS within NS density.
Consequently, the extrapolation of the EOS beyond $n_{\rm c,TOV}$ tends to follow the behavior of the EOS near $n_{\rm c,TOV}$.
On the other hand, the `Data+p($M_{\rm max}$)+pQCD($10n_s$)' constraints include pQCD information. 
This leads to a tendency for the EOS to become softer and more accurate as it incorporates the pQCD boundary conditions.

\subsection{The prospect to validate the pQCD constraints with the  $M_{\rm TOV}$ measurement}
\begin{figure}
    \centering
    \includegraphics[width=0.8\textwidth]{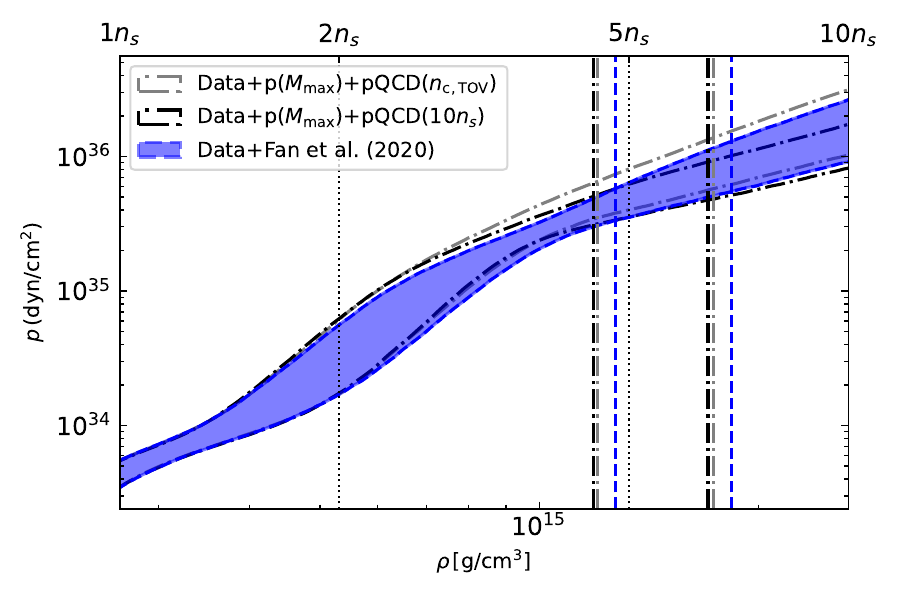}
    \includegraphics[width=0.8\textwidth]{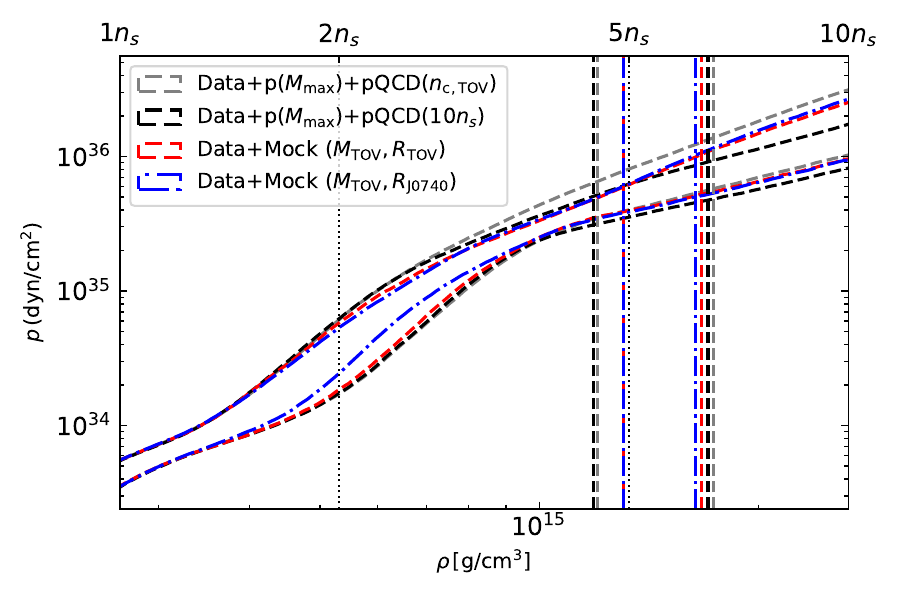}
    \caption{The $90\%$ pressure vs rest-mass density intervals.
    The area filled in blue in the top panel represent the result of the `Data+Fan {\it et al.} (2020)' dataset.
    The gray, black, red dashed lines, and blue dash-dotted lines represent the results of the `Data+p($M_{\rm max}$)+pQCD($n_{\rm c,TOV}$)',  `Data+p($M_{\rm max}$)+pQCD($10n_s$)', `Data+Mock ($M_{\rm TOV}$, $R_{\rm TOV}$)', and `Data+Mock ($M_{\rm TOV}$, $R_{\rm J0740}$)' datasets, respectively.
    The results of `Data+p($M_{\rm max}$)+pQCD($n_{\rm c,TOV}$)' and `Data+p($M_{\rm max}$)+pQCD($10n_s$)' datasets are taken from \citet{Fan:2023spm}.
    The vertical lines indicate the intervals of the central density of the maximum-mass nonrotating NS.}
    \label{fig:eos-range-mock2}
    \hfill
\end{figure}

Recent advancements in {\it ab initio} QCD calculations have garnered significant attention for their role in determining the NS EOS \citep{2022PhRvL.128t2701K, 2022arXiv221111414K,  2023ApJ...950..107G, Fan:2023spm, 2023SciBu..68..913H, 2023PhRvD.108i4014B, 2023PhRvD.107a4011B, 2023arXiv230902345M, 2023PhRvC.107e2801S, 2023arXiv230711125Z}.
Nonetheless, the density at which pQCD constraints should be applied continues to be a subject of contention (see e.g., Refs.~\citep{2023PhRvC.107e2801S, 2023ApJ...950..107G, 2023PhRvD.108d3013E, 2023arXiv230711125Z}).
\citet{Fan:2023spm} argued that accepting the EOSs derived exclusively from pQCD constraints at $n_{\rm c,TOV}$ may lead to inconsistencies.
Specifically, when these EOSs are further constrained to comply with pQCD at densities up to $10n_s$, they could exhibit an anomalously low sound speed within the density interval from $n_{\rm c,TOV}$ to $10n_s$.
Such an abrupt reduction in sound speed may not be physically tenable (see Ref.~\citep{Fan:2023spm} for the details).
This perspective gains further support from the $M_{\rm TOV}=2.17_{-0.12}^{+0.15}M_\odot$ ($90\%$ credible interval) inferred from multimessenger observations of GW170817/GRB 170817A/AT2017gfo (see the blue dashed line in Fig.~3 of Ref.~\citep{2020ApJ...904..119F}), as illustrated in the top panel of Fig.~\ref{fig:eos-range-mock2}, where we can find that the $p-\rho$ intervals constrained by `Data+Fan {\it et al.} (2020)' dataset are more consistent with those constrained by `Data+p($M_{\rm max}$)+pQCD($10n_s$)'.
Additionally, as depicted in the bottom panel of Fig.~\ref{fig:eos-range-mock1}, we find that future mock $M_{\rm TOV}$ and $R_{\rm TOV}$ with measurement uncertainties based on prospects from \citet{2024ApJ...960...67T} have the potential to robustly determine whether the pQCD constraints imposed at $10n_s$ are reliable or not.
In the bottom of Fig.~\ref{fig:eos-range-mock2}, we further examine the impact of augmenting the `Data+Mock ($M_{\rm J0740}$, $R_{\rm J0740}$)' dataset with a less precise determination of $M_{\rm TOV}$ (attainable from future multimessenger observations of BNS mergers) on probing the high-density EOS.
Our results indicate that incorporating a $\pm0.04M_\odot$ estimate of $M_{\rm TOV}$ notably enhances the constraining power on the EOS.
These findings highlight the sensitivity of $M_{\rm TOV}$ to the high-density EOS, underlining the importance of precise determination of $M_{\rm TOV}$.

\section{Summary and Discussion} \label{sec:summary}
In this study, we have explored the implications of the coincidence of GW190425 and FRB 20190425A.
First, if we accept that this coincidence led to the formation of an SMNS for GW190425, the determined maximum mass of a nonrotating NS appears to be too large, with $M_{\rm TOV}=2.77_{-0.11}^{+0.17}M_{\odot}$.
This is in tension with those derived from pQCD constraints and multimessenger observations of GW170817.
If this is indeed the case, the secondary object in GW190814 could be a nonrotating NS.
Meanwhile, the reconstructed EOSs significantly differ from those constrained by pQCD, appearing very stiff with the speed of sound displaying a rapid increase.
However, if we solely rely on current mass-radius measurements of NSs, it is challenging to confirm whether such a coincidence is genuine.
Therefore, we have conducted EOS constraints with mock measurements of massive NSs to determine if this can resolve the issue.
We found that if there is a $2\%$ relative uncertainty for the radius of PSR J0740+6620-like NS, it becomes highly probable to clarify whether a very high $M_{\rm TOV}$ is realistic.
We have also examined the impact of mock $M_{\rm TOV}$ and $R_{\rm TOV}$ measurements, considering the measurement accuracy prospects presented in \citet{2024ApJ...960...67T}, in constraining high-density EOSs. 
We found that a PSR J0740+6620-like NS is not sufficiently massive to effectively constrain the EOS at densities approaching the central density of the maximum mass configuration, $n_{\rm c,TOV}$.
However, precise measurements of the maximum mass, $M_{\rm TOV}$, and/or the corresponding radius, $R_{\rm TOV}$, can significantly refine these constraints.
Therefore, the accurate determination of $M_{\rm TOV}$ is crucial for validating the use of pQCD constraints at densities up to $10n_s$.
This highlights the importance of reliably ascertaining $M_{\rm TOV}$ to advance our understanding of very dense matter.

In the near future, the NICER team is anticipated to release measurements of the radius of PSR J0437-4715 with a projected precision of approximately $\pm5\%$ \citep{2019ApJ...887L..27G}.
In parallel, the Five-hundred-meter Aperture Spherical radio Telescope (FAST) in China is expected to enhance the detection of new pulsars substantially.
The construction of the Square Kilometre Array (SKA) and the next-generation Very Large Array (ngVLA) is planned to proceed in phases.
These developments, along with the anticipated launches of the Enhanced X-ray Timing and Polarization satellite (eXTP) \citep{2019SCPMA..6229502Z} and the Spectroscopic Time-Resolving Observatory for Broadband Energy X-rays (STROBE-X) \citep{2018SPIE10699E..19R} in the 2030s, present a promising future for NS physics research, especially in understanding the EOS.
Polarization measurements from eXTP are projected to refine the constraints on the geometric configurations of thermal hot spots on NS surfaces, effectively reducing the mass-radius degeneracy in pulsar profile models \citep{2019SCPMA..6229503W}.
The wide fields of view of both eXTP and STROBE-X will enhance photon collection capabilities, permitting the study of a broader spectrum of NS sources.
This includes those with lower luminosity or slower rotational velocities, facilitating more accurate mass and radius measurements\citep{2019AIPC.2127b0008W}.
Therefore, the $2\%$ measurement uncertainty assumed in this study is a realistic expectation for NSs with precisely known masses.
On the other hand, the LIGO/Virgo/KAGRA collaboration has initiated their fourth observation run (O4), during which the detectors are expected to progressively reach the projected sensitivity levels.
With the onset of the O5 run, the detectors are scheduled for enhancement to the A+ sensitivity level \citep{2020LRR....23....3A}.
Consequently, the detection horizon for O4 is projected to surpass that of the previous observation run (O3) by $50\sim90\%$, and for the subsequent O5 run, the range may potentially triple that of O3.
Moreover, the integration of emerging detectors such as KAGRA \citep{2021PTEP.2021eA103A} and LIGO-India \citep{2022CQGra..39b5004S} is set to further extend the global network for GW detection.
By the 2030s, it is anticipated that next-generation ground-based GW detectors, including LIGO-Voyager \citep{2020CQGra..37p5003A}, the Einstein Telescope \citep{2010CQGra..27s4002P}, and the Cosmic Explorer \citep{2017CQGra..34d4001A}, will become operational, boasting sensitivities that exceed those of current detectors by at least an order of magnitude.
The expected detection of additional BNS mergers and their electromagnetic counterparts, particularly if an event similar to GW170817 were to occur at a closer distance, holds the promise of a high signal-to-noise ratio for the GW signal.
Concurrent electromagnetic observations are expected to yield more comprehensive data.
These advancements are poised to significantly enhance the precision of tidal deformability measurements and enable the imposition of more stringent limits on the maximum mass of NSs.
Additionally, the augmented sensitivity in high-frequency ranges promises to improve the detection and analysis of GW signals from the post-merger phase which may provide important information to high-density EOS \citep{2022PhRvD.105j4019W, 2022PhRvL.129r1101H, 2023arXiv231006025P}.

\begin{acknowledgments}
This work is supported by the National Natural Science Foundation of China under Grants No. 12233011, No. 12303056, No. 11921003, No. 11933010, and No. 12073080, the Project for Special Research Assistant and the Project for Young Scientists in Basic Research (Grant No. YSBR-088) of the Chinese Academy of Sciences, and the General Fund (Grants No. 2023M733735 and No. 2023M733736) of the China Postdoctoral Science Foundation.
\end{acknowledgments}

\bibliographystyle{apsrev4-1}
\bibliography{ms}

\end{document}